\documentclass[showpacs,preprintnumbers,amsmath,amssymb,preprint,prl]{revtex4}%
\usepackage{amsfonts}
\usepackage{amsmath}
\usepackage{amssymb}
\usepackage{graphicx}%
\begin{document}
\title[CLBs]{\textbf{Cavity Light Bullets: 3D Localized Structures in a Nonlinear Optical
Resonator }}
\author{Massimo Brambilla\footnote{Corrisponding author.
e-mail:brambilla@fisica.uniba.it}, Lorenzo Columbo, Tommaso Maggipinto}
\affiliation{INFM, Dipartimento di Fisica Interateneo, Universit\`{a} e Politecnico di
Bari, via Orabona 4, 70126 Bari, Italy \ }

\begin{abstract}
{\small We consider the paraxial model for a nonlinear resonator with a
saturable absorber beyond the mean-field limit and develop a method to study
the modulational instabilities\textbf{ }leading to pattern formation in all
three spatial dimensions. For achievable parametric domains we observe total
radiation confinement and the formation of 3D localised bright structures. At
difference from freely propagating light bullets, here the self-organization
proceeds from the resonator feedback, combined with diffraction and
nonlinearity. Such "cavity" light bullets can be independently excited and
erased by appropriate pulses, and once created, they endlessly travel the
cavity roundtrip. Also, the pulses can shift in the transverse direction,
following external field gradients.}

\end{abstract}
\date{today}
\maketitle

The competition between transverse diffraction and nonlinearities in optical
systems (resonators, systems with feedback mirror or counterpropagating beams)
leads to the formation of cavity solitons (CS) appearing as bright pulses on a
homogeneous background \cite{opf, ramazza}. The possibility of exploiting CS
as self-organized micropixels in semiconductor vertical microresonators has
been recently proved \cite{nature}, after previous observations and
predictions in various classes of macroscopic optical systems (see e.g.
\cite{ackemann, weiss}). Such phenomena occur when the mean-field-limit (MFL)
is valid, so that in the propagation direction $z$ (coinciding with the
resonator axis) there is no spatial modulation of the field envelope. In this
work we show how, beyond the MFL, an analogous formation of soliton-like
structures confined in the transverse plane \textit{and} in the longitudinal
direction can be predicted. We call these structures Cavity Light Bullets
(CLBs). They appear as bright stable pulses, spontaneously stemming and
self-organizing from the modulational instability (MI) of a homogeneous field
profile, and travelling along the resonator with a definite period.

In the past, the temporal dynamics of a laser field, linked to the competition
of longitudinal modes and without the MFL, has been extensively studied
\cite{ikeda, squicciarini}. Those approaches included a plane wave
approximation and could not account for pattern formation. On the other hand
there are relatively few works dealing with pattern formation beyond the MFL
in an optical resonator. An exact treatment has been proposed in
\cite{LeBerre1} but negligible field absorption is considered, i.e. the field
intensity is assumed constant along the $z$-axis. Other approaches proceed
from a model proposed in \cite{silberberg}, where 2nd-order dispersion is the
main mechanism for longitudinal self-modulation of propagating pulses
\textbf{, }and the adaptation to resonator systems is heuristically provided
with a formalism similar to MFL models . There, the formation of light bullets
is shown \cite{tlidi1, rosanov}. Other theoretical studies have considered
parametric resonators ($\chi^{2}$ nonlinearities) where, though, the mechanism
of structure confinement is due to domain wall locking \cite{oppo}\textbf{
}rather than pattern localization, and experimental confirmations are still
missing. \ \ 

On the other side, the self-collapse of a pulse, leading to formation of light
bullets freely propagating in a medium, has been extensively studied by
various authors \cite{trapani}. Moreover, in this case, the phenomenon does
not involve a MI in the sense of Ref. \cite{firth2} or the localization of
patterns emerging thereof: the formation of one or more bullets involves a
single pulse undergoing diffraction and dispersion. The CLBs presented in our
work, conversely, are stable solutions of the intracavity field profile,
superimposed to a nonmodulated background.

In this Letter, we analyze a well-established model for a resonator with a
saturable absorber \cite{lugiato} to which diffraction in the transverse plane
is added \cite{lax}\textbf{.} The omission of the atomic dynamics, at
difference from \cite{LeBerre1}, allows us to simplify the spatiotemporal
field behavior avoiding competition between different timescales (field and
atoms) \cite{note}.

\textbf{ }After casting the model, we extend an approach developed in the 80s
\cite{squicciarini} to include diffraction and analyze the MI domains and the
pattern formation stemming thereof. We apply checkpoints using existing
literature both close and far from MFL conditions. We then find conditions for
radiation confinement and report about Cavity Light Bullets in a parametric
domain not far (apart from the exclusion of the MFL) from \cite{firth2}.
Finally, we show how CLBs can be excited and their motion controlled via field gradients.

In the paraxial and Slowly Varying Envelope Approximation (SVEA), but without
any hypothesis on the longitudinal profile of the intracavity field, the
equations governing the spatiotemporal dynamics of a two level saturable
absorber in a nonlinear unidirectional ring resonator driven by a coherent
field, can be cast as done in \cite{lugiato1} and after adiabatic elimination
of the atomic variables (good cavity limit) we have%
\begin{equation}
\dfrac{1}{c}\dfrac{\partial F}{\partial t}+\dfrac{\partial F}{\partial
z}=\frac{-\alpha F\left(  1-i\Delta\right)  }{1+\Delta^{2}+|F|^{2}}+\frac
{i}{2k_{0}}\nabla_{\perp}^{2}F \label{eqdinam1}%
\end{equation}
(and complex conjugate) with the boundary condition%
\begin{equation}
F(x,y,0,t)=TY_{inj}+RF(x,y,L,t)e^{-i\delta_{0}} \label{COND1}%
\end{equation}
where $Y_{inj}$, $F$ denote respectively the normalized envelope of the input
beam (CW and transversely homogeneous) and of the intracavity field; $\Delta$
is the scaled\textbf{ } atomic detuning, $\delta_{0}$ is the scaled\textbf{ }
cavity detuning, $\alpha$ is the absorption coefficient per unit length at
resonance. $T$ and $R$ are the transmission and reflection coefficients,
respectively, of the entrance and the exit mirror, being $T+R=1$; $x$ and $y$
are the transverse Cartesian coordinates and $z$ the longitudinal coordinate,
being $z=0$ and $z=L$ the input and the output mirrors, respectively. Here,
for simplicity, we consider the equivalence between the cavity ($\mathcal{L}%
$)\ and the medium's ($L$) lengths, i.e. the active medium completely fills
the cavity.

The transverse laplacian $\nabla_{\perp}^{2}$ describes diffraction in
transverse plane $(x,y)$, while ${\partial F} / {\partial z} $ accounts for
the variation of the field envelope along the cavity axis.

The stationary, transversely homogeneous solution is obtained by setting
$\nabla_{\perp}^{2}F=0$ and $\partial F/\partial t=0$ in Eq. (\ref{eqdinam1})
and by solving the resulting ODE\textbf{;} it turns out that it can be
multivalued. The stationary intracavity intensity $|F_{st}(z)|^{2}$ is
monotonically decreasing from input to output mirror \cite{ikeda, NARDUCCI}.
Then, extending the approach of Lugiato \textit{et al.} \cite{squicciarini} to
include diffraction, we can study the stability of the solutions
$F_{st}(z)=\rho_{st}(z)e^{i\theta_{st}(z)}$, $(\rho_{st},\theta_{st}%
\in\mathcal{R})$ against perturbations which are modulated along $z$ and in
the $(x,y)$ plane; i.e. we consider a perturbed field $F(x,y,z,t)$ as follows
\begin{equation}
F_{st}(z)+\delta F(x,y,z,t)=F_{st}(z)+\delta f(z)e^{\lambda t}e^{i(k_{x}%
x+k_{y}y)} \label{perturb}%
\end{equation}
with $\delta f(z)=e^{i\theta_{st}}\left(  \delta\rho(z)+i\rho_{st}\delta
\theta(z)\right)  $, (to first order, with $\delta\rho,\delta\theta
\in\mathcal{R}$). $k_{x}$ and $k_{y}$ represent the components of the
transverse modulation in the Fourier space. The different ansatz on the
spatial dependence of the perturbations reflects the symmetry invariance in
the $(x,y)$ plane and the lack thereof along the axis. At difference from the
MFL treatment, due the $z$ dependence of $\rho_{st}$ and $\theta_{st\text{ }}%
$, we cannot obtain a simple algebraic equation for the eigenvalues $\lambda$
after linearizing Eq. (\ref{eqdinam1}) around the imposed solution $F_{st}(z)$
for the small perturbation $\delta F(x,y,z,t)$. As it turns out, we must first
find two linearly independent solutions $r_{1}(z)$, $r_{2}(z)$ and $u_{1}(z)$,
$u_{2}(z)$ for the following system of two coupled 1st order ODEs, as obtained
from insertion of ansatz (\ref{perturb}) into Eq. (\ref{eqdinam1}):%
\begin{align}
\frac{dr}{dX}  &  =r\left(  -\frac{1}{A}+\frac{1}{2X}\right)  +u\left(
-\frac{k_{\perp}^{2}A}{4k_{0}\alpha X}\right) \label{ode1}\\
\frac{du}{dX}  &  =r\left(  \frac{\Delta}{A}+\frac{k_{\perp}^{2}A}%
{4k_{0}\alpha X}\right)  +u\left(  \frac{1}{2X}\right)  \label{ode2}%
\end{align}
where $X(z)=\rho_{st}^{2}(z)$, $r(z)=$ $\delta\rho(z)e^{-\frac{\lambda z}{c}}%
$\textbf{, }$u(z)=$ $\delta\theta(z)\rho_{st}(z)e^{-\frac{\lambda z}{c}}$,
$A=1+\Delta^{2}+X(z)$ and $k_{\perp}^{2}$ $=k_{x}^{2}+k_{y}^{2}$. As the field
must fulfill the constraints imposed by the boundary conditions, once such
solutions are found, the perturbed field $F(x,y,z,t)$ can be evaluated, and
then substituted into Eq. (\ref{COND1}). We rehandle the latter equation along
the lines of \cite{squicciarini} and finally, we obtain a quadratic equation
for $\eta=R\,e^{-\frac{\lambda L}{c}}$
\begin{equation}
\eta^{2}W_{1}+\eta W_{2}+W_{3}=0, \label{etaeqq}%
\end{equation}
where the $W_{i}$'s depend on $\delta_{0}$, $\theta_{st}(z)$ and on
$r_{1,2}(z=0)$, $r_{1,2}(z=L)$, $u_{1,2}(z=0)$, $u_{1,2}(z=L)$. The
eigenvalues can thus be obtained as
\begin{equation}
\lambda_{\pm}=\left(  \frac{c}{L}\right)  \left(  -\ln(\eta_{\pm}%
)+\ln(R)-i2\pi n\right)  \text{, \ }n\in\mathcal{N} \label{lambdapm}%
\end{equation}
where it is evident that the condition $Re(\lambda_{\pm})>0$, allows for a
numerable set of unstable longitudinal modes to destabilize the stationary
solution, for each $(k_{x},k_{y})$ wavevector. From a physical view point we
can then expect a complex unstable dynamical behavior of the system associated
to a nonlinear competition among transverse and longitudinal modes. A detailed
treatment of the stability analysis and of the guidelines for accessing a
parametric domain where localization in $3D$ could be achieved, fall beyond
the letter format: an extended paper is currently in preparation.

We tested this stability analysis by reproducing the instability scenario
predicted in \cite{Stefani}: when the values of $\Delta$, $\alpha L$ and
$\delta_{0}$ approach the mean field limit (according to definition introduced
in \cite{lugiato}), the number of longitudinal modes, setting off the
instability of the stationary solutions,\textbf{ } reduce to one.

Beyond the MFL, we could use existing literature as a checkpoint, to reproduce
some patterns previously reported far from the MFL,\textbf{ }e.g. the
multiconical character of the MI with a parametric set close to that
considered in Patrascu \textit{et al. }\cite{LeBerre1}.

The numerical integration of the dynamical equation Eq.(\ref{eqdinam1}) has
shown in this case the formation of three dimensional patterns strongly
irregular in both space and time; this behavior agrees with the preliminary
analyses suggesting chaotic dynamics and spatial decorrelation of patterns as
the input field intensity increases beyond the MI threshold.

\begin{figure}
\begin{center}
\includegraphics[
height=2.6418in,
width=4.3894in
]%
{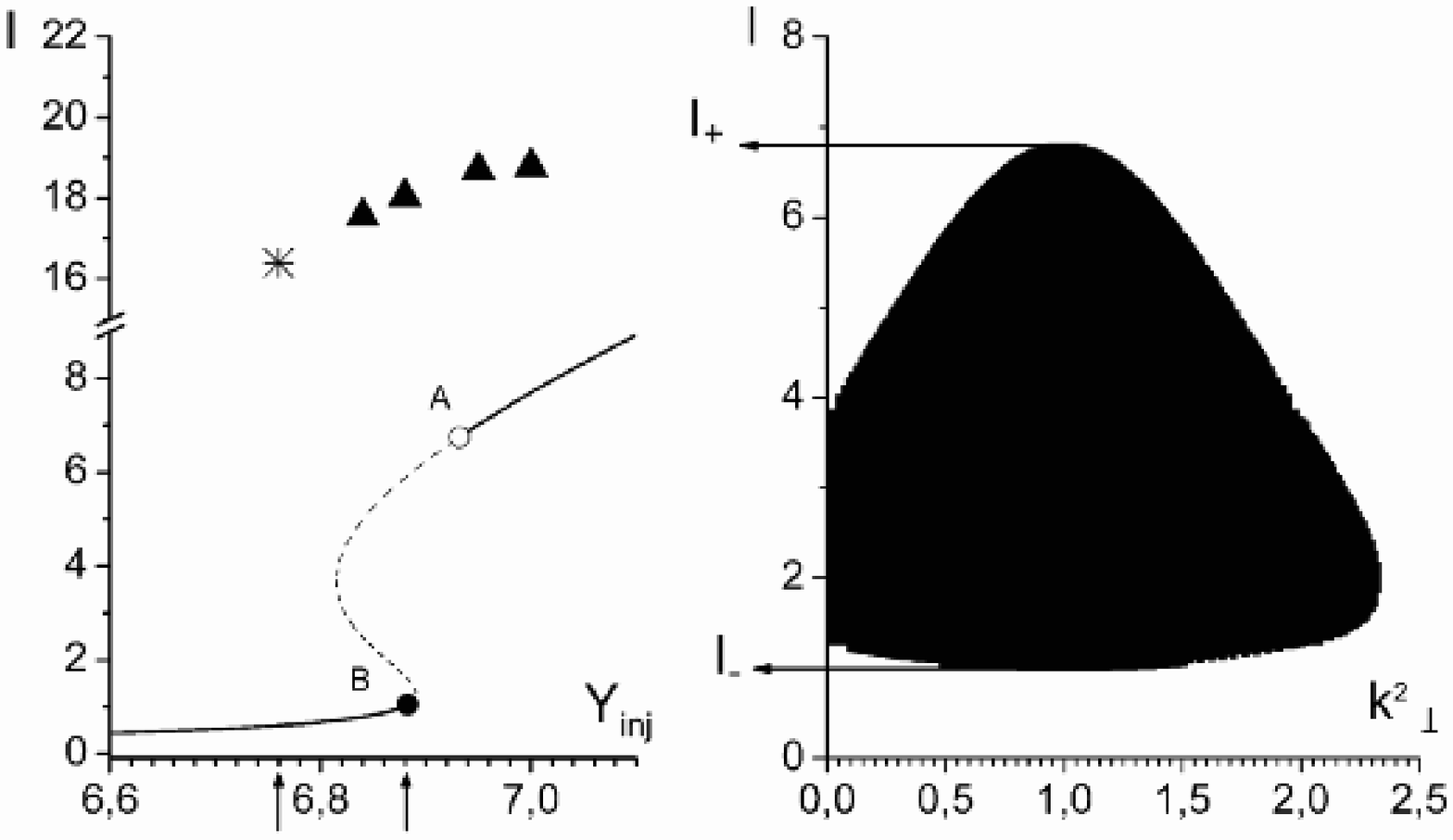}%
\caption{Right: Instability domain. Left: Steady state curve of the
homogeneous solution and results of numerical simulations; continous and
dashed lines refer to stable and unstable portion of the homogeneous branch,
respectively. The ordinate of the asterisk and \ the solid triangles symbols
corrisponds respectively to the mean values of the maximum intensity $I$ of
cavity light bullets and longitudinal rolls and filaments. $I_{-}$ and $I_{+}$
correspond to the minimum and the maximum intensity of the instability region
and hence represent the ordinates of the limiting points $A$ and $B$ ond the
stationary curve. The arrows delimit the interval of \ $Y_{inj}$ where the
modulated solutions coexists with the stationary homogeneous one. }%
\label{instabil}%
\end{center}
\end{figure}

In order to meet stable global patterns and $3D$ localization of light
structures, the path we induced from our analyses was to limit the nonlinear
radiation-matter coupling, without reducing absorption, and also remaining
fairly far from the MFL. A condition we obtained by choosing: $\Delta=0$,
$\alpha L=1.2$, $T=0.1$, $\delta_{0}=0.1$. In Fig. \ref{instabil} we report
for this set of parameters the stationary transversely homogeneous state curve
by plotting the intensity of intracavity field on the exit window
$I=|F(L)|^{2}$ versus the input filed amplitude, and the instability domain
($Re(\lambda_{\pm})>0$) obtained using the stability analysis just described. 

\begin{figure}
\begin{center}
\includegraphics[
height=2.6418in,
width=3.9493in
]%
{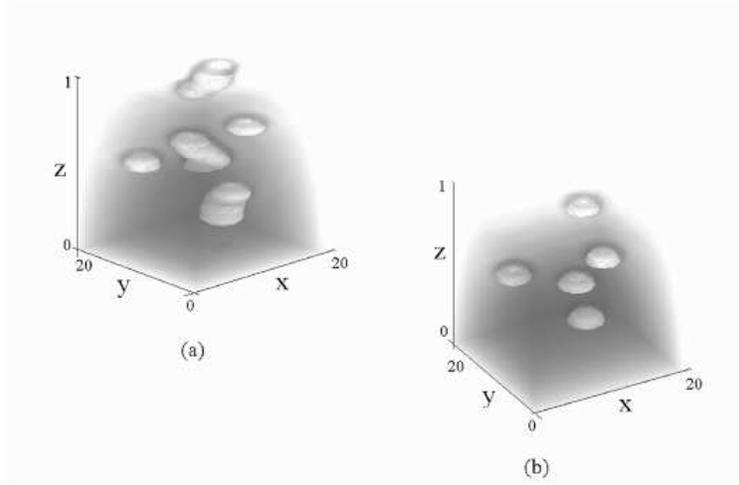}%
\caption{Example of spontaneous self-organization of the system. Gray scale
isosurface of intracavity field for $Y_{inj}=6.74$. $3D$ patterns in the
intensity profile (a) can collapse in localized structures (b) which make a
round cavity trip in a period of $\sim L/c$. Note that in all numerical
simulations the transverse spatial unit is given by the quantity $L/2k_{0}T$,
while the longitudinal variable is scaled to $L$. Finally the time is scaled
to the quantity $L/cT$ so that the interval between the frames represented in
this figure is around $545$ time units.}%
\label{CLB3D}%
\end{center}
\end{figure}

In this case, the simulations show, for example, a transverse intensity
profile of isolated peaks, irregularly distributed in space and oscillating in
time . Seen in $3D$, we identify a series of structures, emerging from a
subcritical bifurcation of the first positive branch of the steady-state curve
(see Fig. \ref{instabil}). They are confined in the transverse plane but also
have a limited, variable length in the $z$ direction and travel the resonator
with a period $\sim\frac{L}{c}$ (see Fig. \ref{CLB3D}(a)). This is thus a
valid example of spontaneous self-organization of a complex optical system in
3D (and time). Moreover, we can observe a spontaneous reduction of the length
of such structures, until a stable (and minimal) limit length is achieved (see
Fig.\ref{CLB3D}(b)). These structures are what we call CLBs.

For input field values where structures coexists with the lower stable
homogeneous branch, delimited in Fig.\ref{instabil} by the two arrows, by
means of the usual technique adopted to switch on/off 2D Cavity Solitons in
the mean field regime \cite{Stefani}, we added to the input field a suitably
narrow and short Gaussian pulse, to locally realize a portion of the spatially
modulated solution.

By varying the Gaussian pulse duration, intensity and phase, we managed to
"write" and "erase" the CLBs, i.e. structures confined in all three spatial
dimensions (analogous to the ones reported in Fig. \ref{CLB3D}) which make a
round-trip in a period $\sim\frac{L}{c}$; when the addressing pulse is much
longer, the length of the structure reaches the full resonator's length and
the stable structure emerging thereof becomes the 3D analog of the 2D cavity soliton.

\begin{figure}
\begin{center}
\includegraphics[
natheight=2.196600in,
natwidth=2.591800in,
height=1.8in,
width=2.6334in
]%
{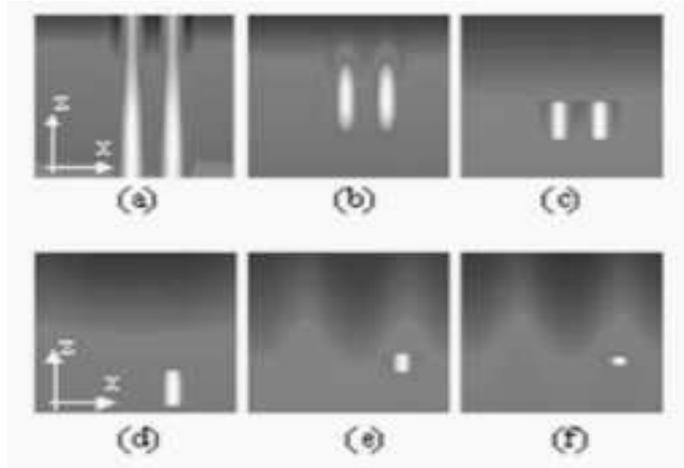}%
\caption{One transverse dimension case. Gray scale plots of intensity
intracavity field profile; white represents the maximum values. (a, b, c)
Sequence of frames showing the writing process of two indipendent CLBs (the
whole sequence is 23 t.u. long); (d, e, f) CLB motion induced via a phase
gradient in the holding beam (the whole sequence is 520 t.u. long). }%
\label{driftmotion}%
\end{center}
\end{figure}

The computational most demanding simulations have been carried on in a
simplified system with only one transverse dimension, $x$, after checking the
coherence of the pattern scenario between the $3D$ and $(z,x)$ systems. We
show that is possible to create several independent CLBs using input Gaussian
pulses aimed at any transverse position, provided they are separated at least
by a critical distance on the order of the CLB diameter. Moreover, by
superimposing to the holding beam a transverse phase modulation, we observe a
slow lateral drift of CLBs climbing the intensity hill, induced by the input
phase gradients themselves, towards their maxima (see Fig. \ref{driftmotion}).

We believe that the lower limit in the self-organised CLB length is linked to
some characteristic length of the system, but we are still missing an
unquestionable analytical result. We report that when longer pulses are
applied in the input field, one seems to achieve longer CLBs (in the $z$
direction); an organized report about the latter issues will be published
elsewhere. The CLB reported in this work can be highly appealing for optical information
processing, making possible the architecture of self-clocked, self-organised,
reconfigurable pixels, which can encode all-optical information both serially
(in the longitudinal trains of CLBs) and parallely (in the transversal arrays
of CLBs). With respect to manipulation of 2D arrays of CS (proposed e.g. in
\cite{spinelli, spinelli2}) one can figure out additional controls (e.g.
transversely injected fields) to change the phase of CLB trains and thus
manipulate information contents,\textbf{ }or to increase the "input" channels,
which can also be seen as logical gates' operands.

On different grounds, CLBs lend themselves to similar applications as usual
light bullets do, namely stroboscopes for atomic/molecular dynamics. Presently
(because of the SVEA in our model) our timescales are still too long for this,
but the benefit unique to cavity-sustained structures is that they can be seen
as a coherent state of the radiation in the cavity that may repeatedly
interact\textbf{ } with (e.g.) a quantum system (atom/condensate). In this
case one could speak of a "quantum stroboscope" provided the coherences in the
system are long compared to the cavity round-trip.

This work was supported by the MIUR National Project "Formazione e controllo
di solitoni di cavit\`{a} in microrisonatori a semiconduttore"

\bigskip
\bibliographystyle{aaai-named}
\bibliography{articles}

\begin{thebibliography}{99}
\bibitem {note}Although the resulting model is not ideally suited to describe
the excitonic response of a semiconductor \cite{brambilla} because of its slow
carrier dynamics, we can note that as far as pattern formation is involved, it
has been shown in literature how the qualitative pattern and soliton scenario
in the MFL saturable absorber \cite{firth2} is not radically different from
the semiconductor case \cite{spinelli}.
\end{thebibliography}

\bigskip%
\pagebreak
%
%

\end{document}